\documentstyle[aps,preprint,epsfig,floats]{revtex}

\tighten

\begin{document}
\def\be{\begin{equation}}
\def\ee{\end{equation}}
\def\bea{\begin{eqnarray}}
\def\eea{\end{eqnarray}}

\draft

\preprint{\parbox{6cm}{\flushright UAB--FT--472
\\September
   1999\\[1cm]}}

\vspace{1in}

\title{\Large \bf Bounds on the Coupling \\ of 
Light Pseudoscalars to Nucleons \\ 
from Optical Laser Experiments\footnote{To be published in Phys. Rev. D}}

\author{ {\bf E. Mass\'o}
\thanks{masso@ifae.es} 
}                   

\address{ Grup de F\'{\i}sica Te\`orica and Institut 
de F\'{\i}sica d'Altes
Energies\\Universitat Aut\`onoma de Barcelona\\ 
08193 Bellaterra, Barcelona, Spain} 

\maketitle

\vfill

\begin{abstract} 

We find the following improved laboratory bounds on the coupling of light
pseudoscalars to protons and neutrons: $g_p^2/4\pi < 1.7 \times 10^{-9}$
and $g_n^2/4\pi < 6.8 \times 10^{-8}$. The limit on $g_p$ arises since a
nonzero $g_p$ would induce a coupling of the pseudoscalar to two photons,
which is limited by experiments studying laser beam propagation in
magnetic fields. Combining our bound on $g_p$ with a recent analysis of
Fischbach and Krause on two-pseudoscalar exchange potentials and
experiments testing  the equivalence principle, we obtain our limit on
$g_n$.
 
\end{abstract}                                                                



The recent work of Fischbach and Krause in references
\cite{FK1} and \cite{FK2} has reopened the issue of the laboratory
constraints on the Yukawa couplings
$g$ of a light pseudoscalar to fermions, defined through the Lagrangian
density
\be
{\cal L}_Y = i g\, \bar \psi(x) \gamma_5 \psi(x)\, \phi(x) \ \ ,
\ee
that couples the pseudoscalar field $\phi(x)$ to the fermion field
$\psi(x)$.

It is well known that the exchange of a light $\phi$ leads to a
spin-dependent long-range interaction among fermions. For fermions
separated by a distance $r$, in the limit that the pseudoscalar mass 
$m \ll 1/r$, the potential is
\be\label{Vone}
V^{(2)}= \frac{g^2}{16\pi} \frac{1}{M^2} \frac{S_{12}}{r^3} \ \ .
\ee
(We display the formula for the particular case of identical fermions of
mass $M$.) The spin-dependent factor $S_{12}$ of the potential 
(\ref{Vone}) reads
\be
S_{12}
= 3 \frac{(\vec \sigma_1 \cdot \vec r) (\vec \sigma_2 \cdot \vec r)}{r^2}
- (\vec \sigma_1 \cdot \vec \sigma_2) \ \ ,
\ee
with $\vec \sigma_i/2$ ($i=1,2$) the spins of the two fermions. The
laboratory experiments trying to constrain such spin-dependent 
interaction lead to relatively poor bounds on the Yukawa couplings 
$g$ to fermions.

In \cite{FK1,FK2} the authors have noticed that a significant
improvement on the bounds on $g$ for nucleons can be obtained by 
considering the potential arising from two-pseudoscalar exchange,
\be\label{Vtwo}
V^{(4)}= - \frac{g^4}{64 \pi^3}  \frac{1}{M^2} \frac{1}{r^3}
\ee
where again we took the $m \rightarrow 0$ limit and the particular 
case of two identical fermions.

When comparing the potentials (\ref{Vone}) and (\ref{Vtwo}) one may think
that to consider $V^{(4)}$  would lead to worse bounds since it has a
$g^2/4\pi^2$ suppression relative to the potential
$V^{(2)}$. However, $V^{(4)}$ is spin independent and thus it is
constrained by experimental searches for such new macroscopic forces.
Fischbach and Krause have shown that the second effect dominates over the
first one when considering Yukawa couplings $g_p$ to protons and $g_n$ to
neutrons. In \cite{FK1}, these authors use data from experiments
testing the equivalence principle \cite{EP}. In \cite{FK2}, they
use data from experiments testing the gravitational inverse square law
\cite{IS}. From the combination of both types of limits they 
finally get \cite{FK2}
\bea
\frac{g_n^2}{4 \pi} &<& 1.6 \times 10^{-7} \  \ ,
\\
\frac{g_p^2}{4 \pi} &<& 1.6 \times 10^{-7} \  \ .
\eea

In the present short paper, we would like to show that there are further
laboratory constraints on the Yukawa couplings $g_n$ and $g_p$. We 
consider the coupling of $\phi$ to two photons induced by the triangle
diagram that we display in the figure. In the loop, the internal line
is a proton since it couples to the pseudoscalar and to the photons. As
expected, the evaluation of the triangle diagram leads to a
gauge-invariant effective Lagrangian density of the form
\be
{\cal L}_{\phi\gamma\gamma}= \frac{1}{8}\, f\, 
\epsilon_{\mu \nu \alpha \beta}
F^{\mu \nu}(x) F^{\alpha \beta}(x)\, \phi(x)
  \ \ ,
\ee
with $F^{\mu \nu}(x)$ the photon field strength. One gets \cite{anomaly}
\be \label{relationship}
f= \frac{\alpha}{\pi M_p}\, g_p  \  \ .
\ee
($M_p$ is the proton mass).

\begin{figure}[bht]

\begin{center}

\epsfig{file=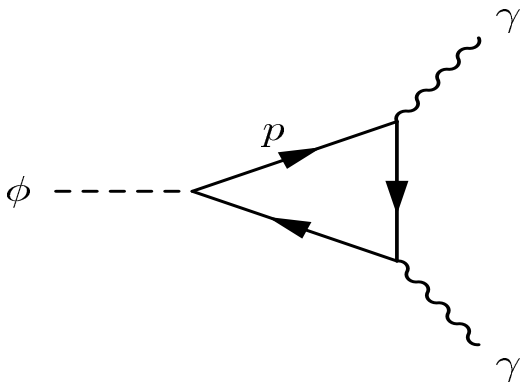,width=7cm,height=5cm}

\end{center}

\caption{{\it One of the two triangle diagrams inducing the effective 
$\phi\gamma\gamma$ interaction. The other diagram is the crossed one.}}

\end{figure}

The $\phi\gamma\gamma$ coupling $f$ is suppressed by a factor $\alpha$
compared to the Yukawa coupling $g_p$ but, as shown below, the
existing laboratory constraints on $f$ allow us to place a stringent 
bound on $g_p$.

Among all the laboratory limits on $f$ \cite{MT} the most restrictive
ones come from the study of laser beam propagation through a transverse
magnetic field. A light pseudoscalar coupled to two photons would induce
effects such as optical rotation of the beam polarization, the 
appearance of ellipticity of the beam, and photon
regeneration \cite{cameron}. The absence of these effects in the data 
of the experiment leads to the limit 
\cite{cameron} 
\be \label{laser} 
f < 3.6 \times 10^{-7}\, {\rm GeV}^{-1}  \ \ .
\ee
Using now the relationship in Eq.(\ref{relationship}), the previous 
limit translates into a bound on the proton Yukawa coupling 
\be \label{limit_p}
\frac{g_p^2}{4 \pi} < 1.7 \times 10^{-9} \ \ .
\ee
The limit (\ref{laser}) is valid for masses of the light pseudoscalar
$m<10^{-3}$ eV. It follows that our bound (\ref{limit_p}) is also valid
for this mass range. This corresponds to interaction ranges of the
potentials (\ref{Vone}) and (\ref{Vtwo}) larger than about $0.02$ cm. 

We can now constrain the neutron Yukawa coupling by combining our
bound (\ref{limit_p}) with the results from \cite{FK1}. As 
explained above, constraints on $g_n$ and $g_p$ can be placed
by considering the $V^{(4)}$ potential. In reference \cite{FK1}, 
the implications for the couplings $g_n$ and $g_p$ from the equivalence
principle experiment \cite{EP} have been  worked out in detail. The
final result is a constraint on a combination of both Yukawa
couplings \cite{FK1}
\be \label{gundlach}
(9.6 g_p^2 + 15.3 g_n^2)|0.05925 g_p^2 - 0.05830 g_n^2| \\
< 6.4 \times 10^{-13} \ \ .
\ee
Introducing (\ref{limit_p}) in (\ref{gundlach}) we find the stringent 
bound
\be \label{limit_n}
\frac{g_n^2}{4 \pi} < 6.8 \times 10^{-8} \ \ .
\ee

The ongoing experiment PVLAS \cite{PVLAS} that also studies laser beam
propagation is supposed to ameliorate the present limit (\ref{laser}) on
$f$. This will in turn improve our bounds (\ref{limit_p}) and
(\ref{limit_n}). 

In summary, we have shown first that the stringent bound in 
Eq.(\ref{limit_p}) on the coupling $g_p$ of the proton to a light 
pseudoscalar (with $m<10^{-3}$ eV) can be obtained by considering the
induced coupling of the pseudoscalar to two photons which in turn is
limited by laser propagation experiments. Second, we have combined our
bound on $g_p$ with the results coming from data on equivalence principle
experiments constraining the spin-independent potential due to
two-pseudoscalar exchange. As a result, we are able to put the stringent
bound in Eq. (\ref{limit_n}) on the neutron coupling $g_n$ to the light
pseudoscalar.

\acknowledgments

Work partially supported by the CICYT Research Project
AEN98-1116. We would like to thank Francesc Ferrer for helpful
discussions.

\end{document}